\documentclass[reprint,superscriptaddress]{revtex4-1}

\pdfoutput=1
\usepackage[utf8]{inputenc}

\usepackage{mathtools}
\usepackage{amsmath,amsthm,enumerate,fancyhdr,titlesec,framed}

\usepackage{natbib}
\usepackage{subcaption}

\usepackage[bitstream-charter]{mathdesign}
\usepackage{hyperref}
\usepackage{qcircuit}
\begin{document}

\title{Generating 3 qubit quantum circuits with neural networks}

\author{Michael Swaddle}
\email{michael.swaddle@research.uwa.edu.au}
\affiliation{School of Physics, The University of Western Australia, Crawley 6009, Australia}

\author{Lyle Noakes}
\affiliation{Faculty of Engineering, Computing and Mathematics, The University of Western Australia, Crawley 6009, Australia}

\author{Liam Salter}
\author{Harry Smallbone}
\affiliation{The University of Western Australia, Crawley 6009, Australia}

\author{Jingbo Wang}
\affiliation{School of Physics, The University of Western Australia, Crawley 6009, Australia}

\date{\today}

\begin{abstract}
\noindent
    A new method for compiling quantum algorithms is proposed and tested for a three qubit system.  The proposed method is to decompose a a unitary matrix \(U\), into a product of simpler \(U_j\) via a neural network.  These \(U_j\) can then be decomposed into product of known quantum gates.  Key to the effectiveness of this approach is  the restriction of the set of training data generated to paths which approximate minimal normal subRiemannian geodesics, as this removes unnecessary redundancy and ensures the products are unique.  The two neural networks are shown to work effectively, each individually returning low loss values on validation data after relatively short training periods. The two networks are able to return coefficients that are sufficiently close to the true coefficient values to validate this method as an approach for generating quantum circuits. There is scope for more work in scaling this approach for larger quantum systems.
\end{abstract}

\maketitle

\section{Introduction}
    In 1982, Feynman showed that a classical Turing machine would not be able to efficiently simulate quantum mechanical systems \cite{feynman1982simulating}. Feynman went on to propose a model of computation based on quantum mechanics, which would not suffer the same limitations. Feynman's ideas were later refined by Deutsch who proposed a \textit{universal quantum computer} \cite{Deutsch97}. In this scheme, computation is performed by a series of \textit{quantum gates}, which are the quantum analog to classical binary logic gates. A series of  gates is called a \textit{quantum circuit} \cite{nielsen-book}. Quantum gates act on \textit{qubits} which is the quantum analog of a bit.
    \\
    \\
    Lloyd later proved that a quantum computer would be able to simulate any quantum mechanical system efficiently \cite{lloyd1996universal}. Equivalently, this can be stated as; given some special unitary operation \( U \in \mathrm{SU}(2^n)\), \( U^{\dagger} U = I\), there exists some quantum circuit that  approximates \(U\), where \(n\) is the number of qubits.  One pertinent question that remains is how to find the circuit which implements this \(U\). In certain situations the circuit to implement \(U\) can be found exactly. However in general it is a difficult problem, and it is acceptable to approximate \(U\). Previously \(U\) has been found via expensive algebraic means \cite{qcompiler,opt-qcompiler,cosine-sekigawa, Mottonen2004}.  Another novel attempt at finding an approximate \(U\) has been to use the tools of \textit{Riemannian geometry}. 
    \\
    \\
    Nielsen originally proposed calculating special curves called \textit{geodesics} between two points, \(I\) and \(U\) in \(\mathrm{SU}(2^n)\). Geodesics are fixed points of the energy functional \cite{wolfgang}. Nielsen claimed that when an energy minimising geodesic is discretised into a quantum circuit,  this would efficiently simulate \(U\) \cite{nielsen-geom-1,nielsen-geom-2,nielsen-geom-3,nielsen-geom-4,nielsen-geom-5}. In practice however, finding the geodesics is a difficult task. Computing geodesics requires one to solve a boundary value problem in a high dimensional space. Furthermore, Nielsen originally formulated the problem on a Riemannian manifold equipped with a so called \textit{penalty} metric, where the penalty was made large. This complicated solving the boundary value problem \cite{brachistochrone}.
    \\
    \\
    The Nielsen approach can be refined by considering subRiemannian geodesics. A subRiemannian geodesic is only allowed to evolve in directions from a \textit{horizontal subspace} of the tangent space \cite{montgomery}. This approach still involves solving a complicated boundary value problem. For a practical tool, a much faster methodology to synthesise a \(U\) is required. With recent advances in computing power, \textit{neural networks} (NN) are an attractive option.
    \\
    \\
    The problem is to find \(U\) approximately as a product of exponentials
    \begin{align} U \approx \,  &\mathbf{E}(c) = \exp(c^1_1 \tau_{1}) \dots \exp( c^1_m \tau_m ) \nonumber \\
                            &\dots \exp( c^N_1 \tau_1 ) \dots  \exp( c^N_m \tau_{m}), \label{eqn:U} \end{align}
    where \( \mathbf{E} \) we call the \textit{embedding} function, \( c = (c_1^1 ,\dots, c^N_m )\) and the \(\tau_i\) are a basis for a \textit{bracket generating} subset of the Lie algebra \( \Delta \subset  \mathfrak{su}(2^n)\) of dimension \(m\). Bracket generating means that repeated Lie brackets of terms in \(\Delta \) can generate any term in \( \mathfrak{su}(2^n)\). Because products of matrix exponentials generate Lie bracket terms
    \[ \exp(A) \exp(B) = \exp(A+ B+ \frac{1}{2}[A,B] + \dots ),\]
any \(U \in \mathrm{SU}(2^n) \) can be written as Equation (\ref{eqn:U}) with sufficiently many products. We restrict ourselves to \(U\) which can be written as a product of a polynomial in \(n\) terms . An example of such a \(\Delta\) could be the matrix logarithms of universal gates. For convenience it is easier to work with all permutations of Kronecker products of one and two Pauli matrices, so 
    \[ \Delta = \mathrm{span} \{\frac{\mathrm{i}}{\sqrt{2^n}} \sigma_i^j , \frac{\mathrm{i}}{\sqrt{2^n}} \sigma_i^k \sigma_j^l \} ,\]
    where \( \sigma^j_i \) represents the \(N\) fold Kronecker product, \( I \otimes \dots \otimes \sigma_i \otimes \dots \otimes I\), with a \( \sigma_i \) inserted in the \(j\)-th slot and \(I\) representing the \( 2 \times 2 \) identity matrix. Exponentials of these basis elements have very simple circuits, for more detail see Appendix A.
    \\
    \\
    We propose that a neural network be trained to learn \( \mathbf{E}^{-1}\). The neural network will try to find all the coefficients \( c^k_{i} \) so the product approximates \(U\). In this approach, the neural network takes a unitary matrix \(U\) as an input and returns a list \(c\) of \( c^k_{i}\). A segment is a product of \( m \) exponentials of each basis element. In total there are \(N\) segments. We only examine \(U\) which are implementable in a reasonable number of segments. We found that we required two neural networks to achieve this. The first is a Gated Recurrent Unit, GRU, network \cite{gru_paper_1,gru_paper_2} which factors a \(U\) into a product of \(U_j\),
    \[U \approx U_1 U_{2} \dots U_j \dots U_N, \]
    where each \(U_j\) is implementable in polynomially many gates, which we call \textit{global decomposition}.
    The second is simply several dense fully connected layers, which decomposes the \(U_j\) into products of exponentials
    \[ U_j \approx \exp( c^j_1 \tau_1 ) \dots \exp( c^j_m \tau_m ), \]
    which we term \textit{local decomposition}. These procedures can be done with traditional optimisation methods. The lack of a good initial guess meant that it took an order of an hour in \(\mathrm{SU}(8)\). While the output from the neural network may not implement \(U\) to a required tolerance, it does provide a good initial guess as the error will be small. The output from the neural network could be refined with another optimisation algorithm.
\section{Training data \label{training_data}}
    \noindent To generate the training data, the \(c\) should not be chosen randomly. If there is no structure to how \(c\) is chosen, it will introduce extra redundancy. More seriously, \(\mathbf{E}^{-1} \) will not be well defined. There are infinitely many ways to factor a \(U\) into some unordered product of matrix exponentials. Geometrically this could be visualised as taking any path from \(I\) to \(U\) on \(\mathrm{SU}(2^n)\). Randomly generating data may give two different decompositions for a \(U\), and so \(\mathbf{E}\) is not one to one. To ensure the training data is unique, we propose that these paths should be chosen to be, at least approximately, minimal normal subRiemannian geodesics. 
    \\
    \\
    The choice of using geodesics is not particularly special. Other types of curves could be used as long as it uniquely joins \(I\) and \(U\). This is so \( \mathbf{E}^{-1} \) is well defined. Generating random geodesics can be done simply by generating random initial conditions. However the geodesics must also be minimal. The first way to try and ensure they are minimal is to bound the norms of the initial conditions. 
    \\
    \\
    The normal subRiemannian geodesics in \(\mathrm{SU}(2^n)\) can be found via the Pontryagin Maximum Principle \cite{pmp-intro,pmp-book} by minimising the energy functional
    \[ \mathcal{E}[x] = \int_0^1 dt \langle \dot{x}, \dot{x} \rangle, \]
    where \( \langle, \rangle \) is the restriction of the bi-invariant norm to \(\Delta \subset \mathfrak{su}(2^n)\), and \( x :[0,1] \rightarrow \mathrm{SU}(2^n) \). See Chapter 7 of \cite{opt-control-lie-group} for a review. The normal subRiemannian geodesic equations can be written as
    \begin{align*}
        \dot{x} &= u x, \\
        \dot{\Lambda} &= [\Lambda, u ],\\
        u &= \mathrm{proj}_{\Delta}( \Lambda),
    \end{align*}
    where  \( \Lambda : [0,1] \rightarrow \mathfrak{su}(2^n) \),  \( u : [ 0,1] \rightarrow \Delta \subset \mathfrak{su}(2^n) \) and \( \mathrm{proj}_{\Delta} \) is projection onto \( \Delta\).
    This can be re-written as the single equation
    \begin{equation} \dot{x} = \mathrm{proj}_{\Delta}( x \Lambda_0 x^{\dagger} ) x, \label{eqn:geod} \end{equation}
    where \( \Lambda_0 = \Lambda(0) \).
    Choosing the \( \Lambda_0 \) completely determines the geodesic. To generate the training data for the \(U_j\), first randomly choose a \( \Lambda_0\).  The \(U_j \) are then matrices which forward solve the geodesic equations
    \[ x(t_{j+1} ) = U_j x(t_j),\]
    where \( [0,1] \) has been divided into \(N\) segments of width \(h\). For this paper we utilised the simple first order integrator
    \[ U_j = \exp\big( h \, \mathrm{proj}_{\Delta}(x_j \Lambda_0 x_j^{\dagger}) \big),  \]
    since approximating the geodesic is sufficient. There are infinitely many bi-invariant Riemannian geodesics joining \(I \) and \(U\), for the different branches of \( \log(U) \). SubRiemannian geodesics are similarly behaved, but it varies on the norm of \( \Lambda_0\). To generate the training data we bounded the norms by \(  \mathrm{dim}(\Delta) = \mathcal{O}(n^2) \), to try and ensure the geodesics are unique. 
    \\
    \\
    Further, the norm \( || \mathrm{proj}_{\Delta} (\Lambda_0 ) || = || u_0 ||\) determines the distance between \(I \) and a \(U\). Nielsen showed that the distance can be thought of as approximately the complexity to implement \(U\). Lemma (3) in \cite{nielsen-geom-1} shows that a \(U\) further away from \(I\) requires more gates. The distance however is likely to scale exponentially. By bounding the norm by a polynomial, this ensures the training data only contains \(U\) which are reachable with a polynomial number of quantum gates.    
\section{Network Design - SU(8) }    
    \subsection{Global decomposition}
    The neural network for the global decomposition takes an input of \(U\) and returns a list of \(U_j\). To do this \(U\) is decomposed into rows of length \(2^n\). This makes \( 2^n\) real vectors. Each row is treated as a single timestep in the GRU layer. The output \(U_j\) are also decomposed into their rows and these rows are treated as timesteps in the output. This gives \( 2^n N \) output vectors of length \( 2^n\). In particular we examined the \(n=3\) qubit case. For \(\mathrm{SU}(8) \) we found \(10\) stacked GRU layers was sufficient to give reasonable results.  In \(\mathrm{SU}(8)\) we chose \(N=10\) , so there were \( 8 \) input vectors of length \( 8\) and \( 80 \) output vectors of length \( 8\). The network was implemented in the Keras Python library \cite{keras} with the TensorFlow backend, on a Nvidia GTX 1080.
    \subsection{Local decomposition}
    For \(\mathrm{SU}(8)\) a network with \(2\) fully connected dense hidden layers of \(2000\) neurons, with the ReLU activation function was found to be sufficient. The input layer took a vectorised \(U_j\), and outputted \( \dim(\Delta)\) values. The network was implemented in the Keras Python library with the TensorFlow backend, on a Nvidia GTX 1080.
\section{Results - SU(8)}
   \subsection{Global decomposition}
\begin{figure}[h!]
        \centering
        \includegraphics[width=0.45\textwidth]{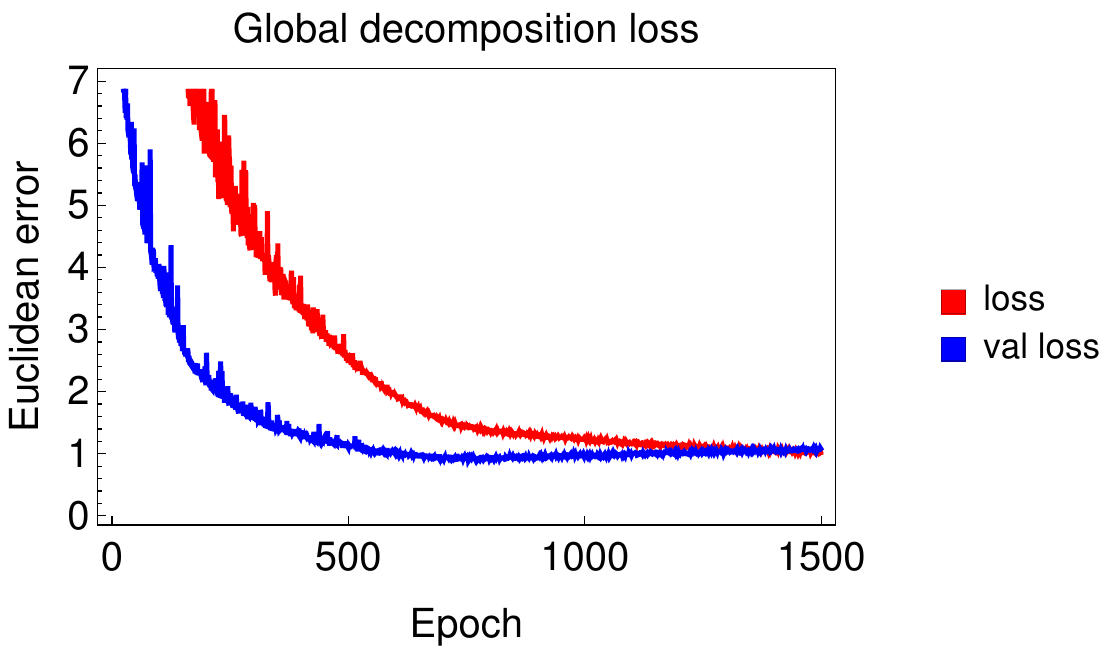}
        \caption{The loss and validation loss from training the global decomposition.}
        \label{fig:gruLoss}
    \end{figure}

\begin{figure}[h!]
        \centering
        \begin{subfigure}{0.48\textwidth}
            \centering
            \includegraphics[width=0.4\linewidth]{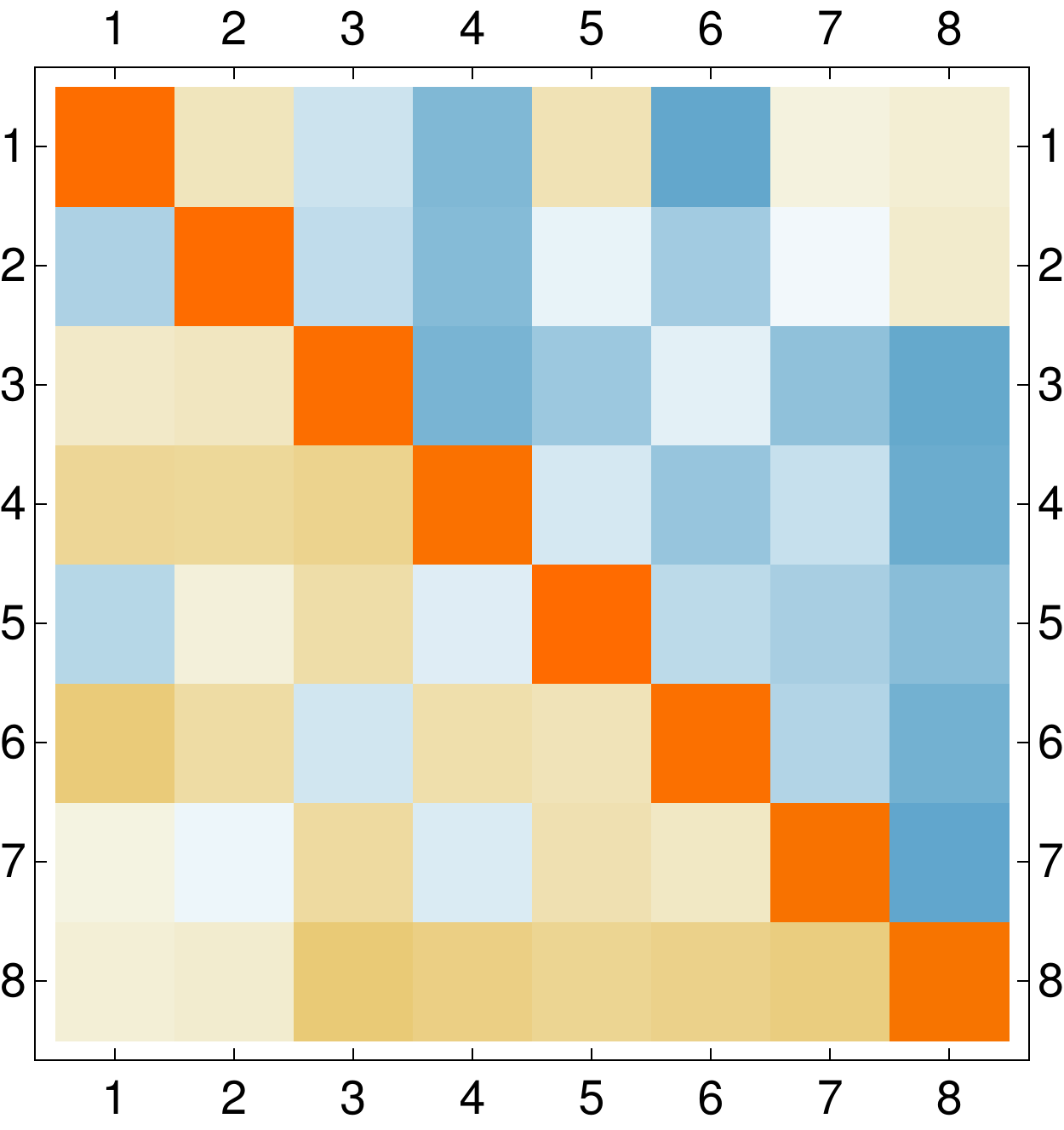}
            \caption{Real components of a \(U_j\) generated by the NN.}
            \label{fig:outUi}
        \end{subfigure}
        \begin{subfigure}{0.48\textwidth}
            \centering
            \includegraphics[width=0.4\linewidth]{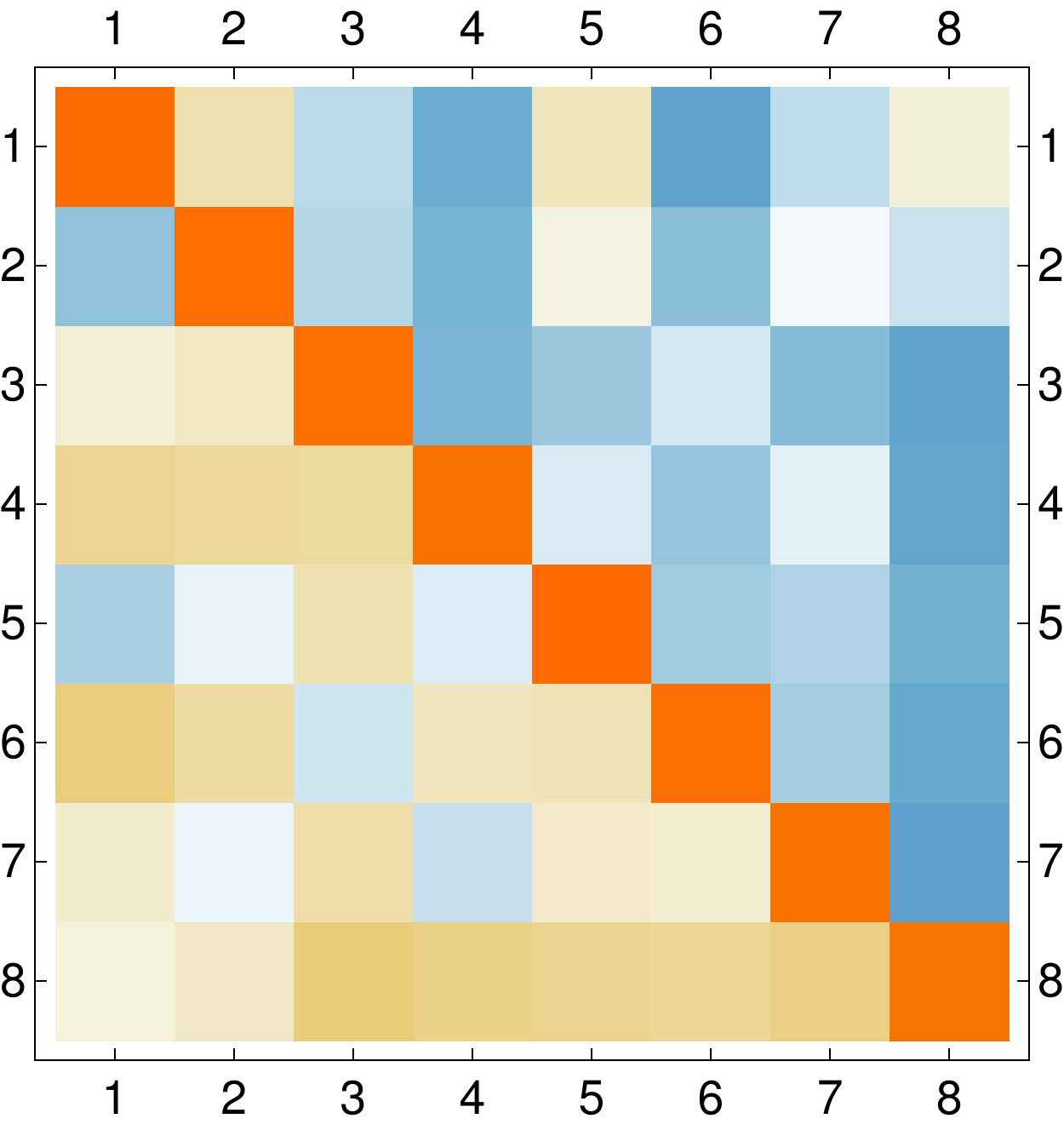}
            \caption{The respective known real components of a \(U_j\) from the validation dataset.}
            \label{fig:validUi}
        \end{subfigure}
        \caption{A known \(U_j\) from the validation data and the \(U_j\) generated by the NN in \(\mathrm{SU}(8)\) for global decomposition. Each \(U_j\) is close to the identity matrix. The shading from blue to orange represents \( [-1,1] \)}
    \end{figure}

\begin{figure}[h!]
        \centering
        \begin{subfigure}{0.48\textwidth}
            \centering
            \includegraphics[width=0.6\linewidth]{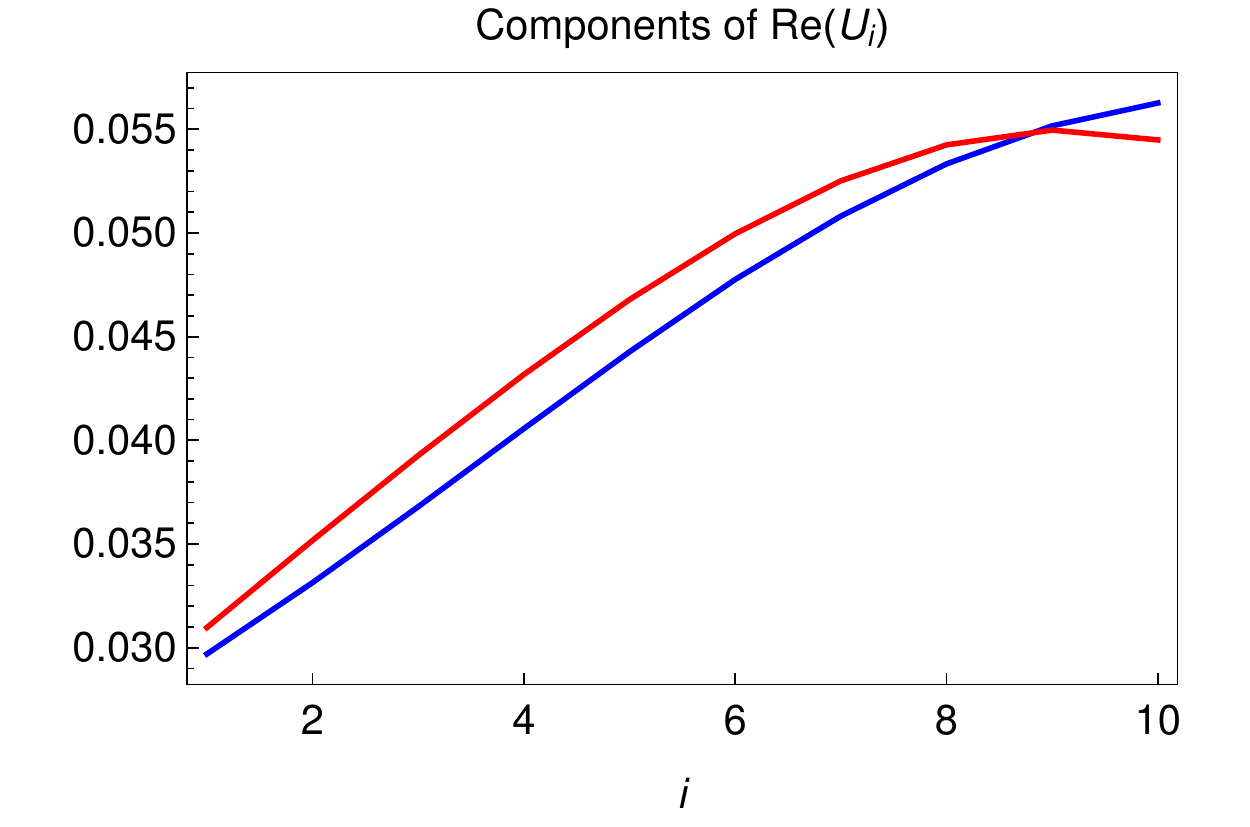}
            \caption{The same real entry from the \(10\) \(U_i\) from the validation data set (blue), vs the predicted output (red).}
            \label{fig:u34}
        \end{subfigure}
        \begin{subfigure}{0.48\textwidth}
            \centering
            \includegraphics[width=0.6\linewidth]{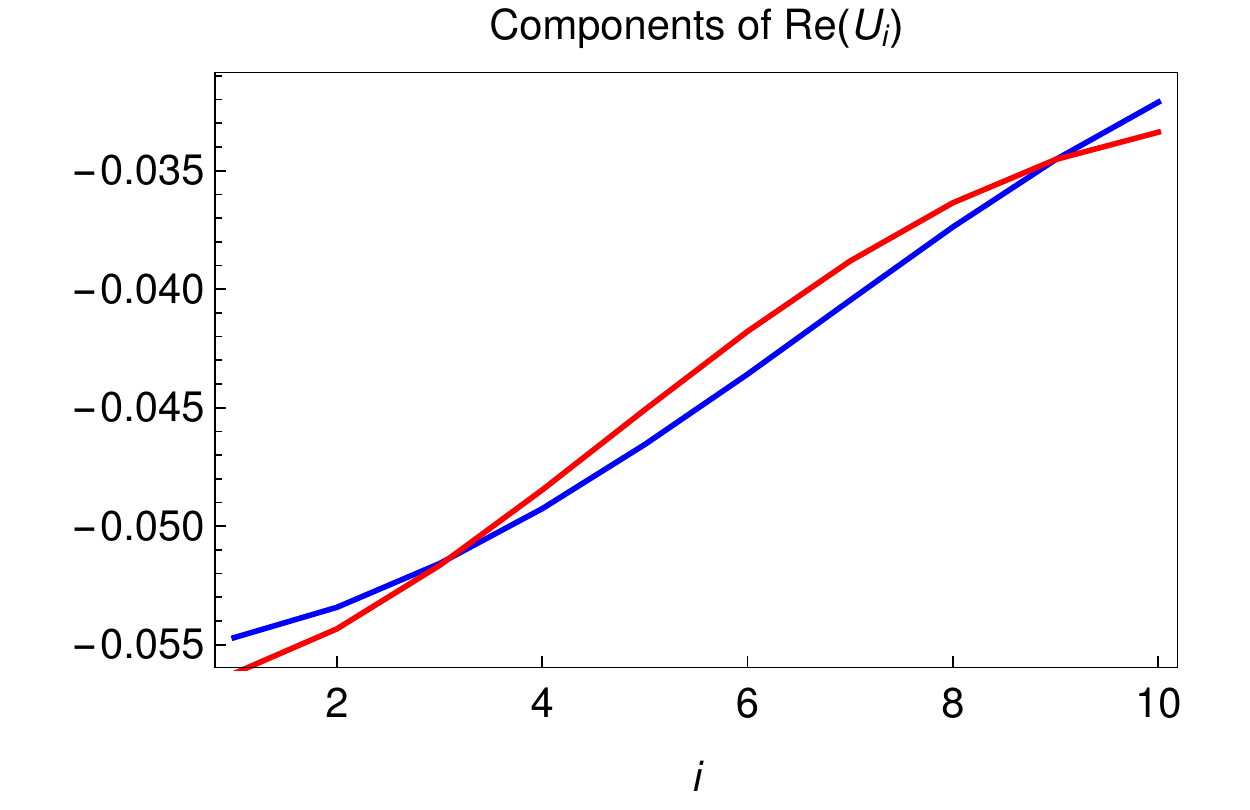}
            \caption{The same real entry from the \(10\) \(U_i\) from the validation data set (blue), vs the predicted output (red). }
            \label{fig:u25}
        \end{subfigure}
        \caption{Real entries of validation \(U_i\) vs the \(U_i\) generated by the NN. Recall the \(U_i\) are not constant, and solve equation (\ref{eqn:geod}). The behaviour displayed here was typical in other entries. }
    \end{figure}

    The global decomposition network was trained on \(U_j\) taken from \(5000\) randomly generated geodesics in \(\mathrm{SU}(8)\). \(500\) were used for validation data. The loss function used was the standard Euclidean distance between the output vector and the desired output. After \( 1500 \) training epochs the validation loss reached \( \sim 0.9 \) and did not decrease. This was found to be sufficient to generate \(U_j\) close to the training data. Figure (\ref{fig:gruLoss}) shows the validation and training loss. Figure (\ref{fig:outUi}) and figure (\ref{fig:validUi}) shows a randomly chosen \( U_j\) from a list of \(U_j\) generated by the network, and from the training data respectively for some random \(U\). Most \(U_j\) appeared to be very similar. Figure (\ref{fig:u34}) and figure (\ref{fig:u25}) show the same entry in consecutive \( U_i\) for validation data. Again the network was able to output values very close to the values in the validation dataset. This similarity was typical. This shows the network is able to reasonably approximate the \(U_j\). 

\subsection{Local decomposition}
\begin{figure}[h!]
        \centering
        \includegraphics[width=0.45\textwidth]{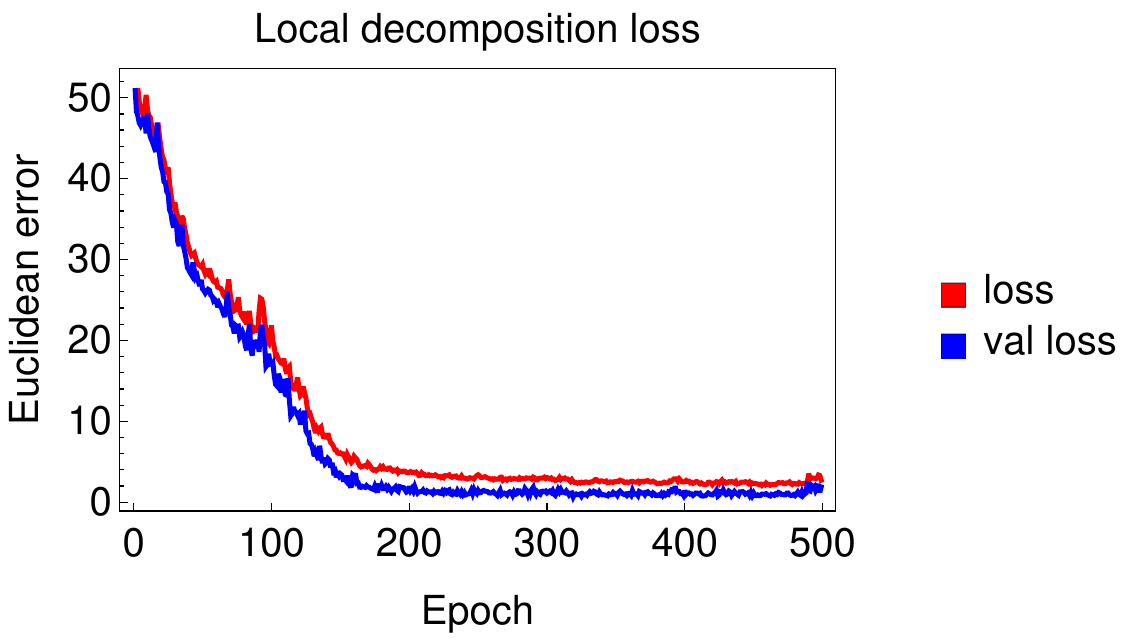}
        \caption{The loss and the validation loss from training the local decomposition. There was no significant improvement after \(500\) epochs.}
        \label{fig:denseLoss}
    \end{figure}

\begin{figure}[h!]
        \centering
        \begin{subfigure}{0.48\textwidth}
            \centering
            \includegraphics[width=0.4\linewidth]{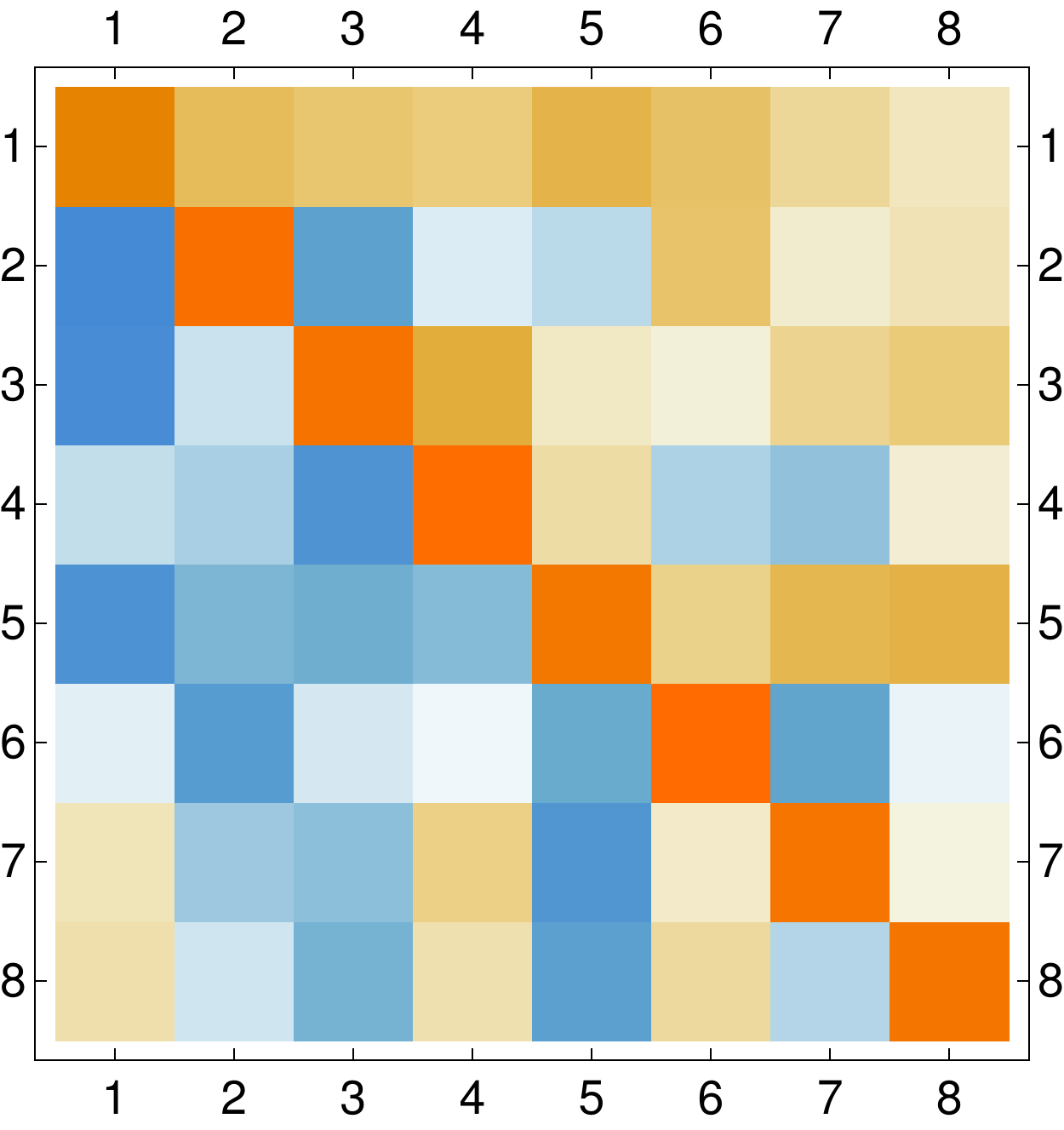}
            \caption{Real components of a \(U_j\) generated by the NN.}
            \label{fig:localoutUi}
        \end{subfigure}
        \begin{subfigure}{0.48\textwidth}
            \centering
            \includegraphics[width=0.4\linewidth]{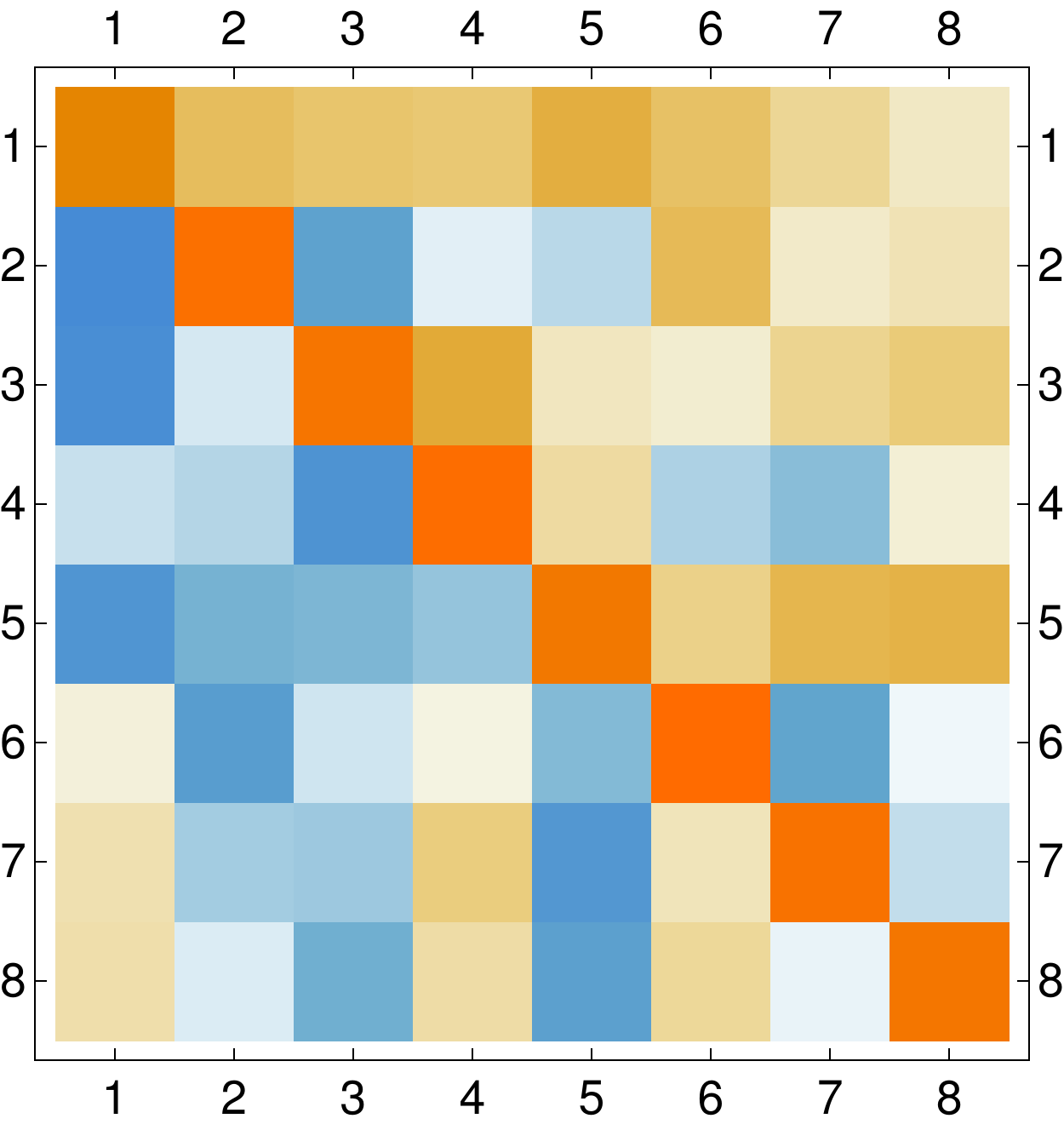}
            \caption{The respective known real components of a \(U_j\) from the validation dataset.}
            \label{fig:localvalidUi}
        \end{subfigure}
        \caption{A known \(U_j\) from the validation data and the \(U_j\) generated by the NN in \(\mathrm{SU}(8)\). These figures are for the local decomposition network. The shading from blue to orange represents \( [-1,1] \)}
    \end{figure}

The network to implement the local decomposition was trained on \(U_j\) generated by choosing a random \(m\)-vector of the coefficients \(c^j_i\), where each \(c^j_i \) was order \( 1/N \). In total there were \(5000\) pairs in the training set, and \(500\) in the validation set. Figure (\ref{fig:denseLoss}) shows the validation and training loss. After \( 500\) epochs the network was able to sufficiently compute the local decomposition to reasonable error (on average 0.16).  Figures (\ref{fig:localoutUi}) and (\ref{fig:localvalidUi}) show a matrix generated by the neueral network and the target matrix.

\section{Conclusion}
Training two neural networks to together decompose \(U\) into \(c^j_i \) via a two-step approach (global decomposition followed by local decomposition) was found to be successful, when restricting the set of training data generated to paths which approximate minimal normal subRiemannian geodesics.  This restriction limited the training data pairs to ones which were one-to-one, eliminating redundancy.  For the global decomposition, using a neural network consisting of stacked GRU layers allowed for efficient training of the network, with the validation loss of the network approaching its minimum at 500 epochs for \(\mathrm{SU}(8)\).  A simple dense network with two hidden layers proved sufficient for the local decomposition.  In \(\mathrm{SU}(8)\), the networks were small enough that both networks were able to be trained on a desktop machine with a single NVidia GTX 1080 GPU.  The two stage decomposition proved more successful than single-stage attempts to form a solution, with the decomposition of a given \(U\) into \(U_j\) being crucial for this increase in effectiveness.  This approach to the solution for this problem demonstrates a novel use of neural networks.
\\
\\
Although this approach works well for systems with small numbers of qubits (such as the \(\mathrm{SU}(8)\) case used as an example), the approach does not scale well with increasing number of qubits. This is because the size of the network scales by the number of entries in matrices in \(\mathrm{SU}(2^n)\). Although this is not a significant problem for currently realisable quantum computers, or those in the near future, it will increasingly become problematic as quantum computing continues to advance.  To somewhat counteract this, the complexity of the problem can be decreased by restricting the set of \(U\) on which the neural network is trained. For example if the \(U\) are sparse, some savings in the size of the network may be made. Investigating this will be increasingly significant, as it will increase the practical usefulness of this approach.  
\\
\\
As noted in section \ref{training_data}, the choice of using geodesics to restrict the training data is fairly arbitrary, and as such, there may be different ways of restricting the training data which, while still ensuring the input/output is one-to-one, may produce a better dataset, improving the accuracy of the networks. This is heavily related to the nature of \( \Lambda_0 \) which is currently not fully understood. Exploring this problem is a possible future avenue of investigation, which may improve the effectiveness of the approach described in this paper.  
\\
\\
Finally note that training the network is the most computationally expensive part of this approach. Once the network is trained, propagating an input through through the network is much more efficient than the conventional optimisation techniques for compiling \(U\). 
\\
\\
All data and programs used to produce this work can be found at \href{https://github.com/Swaddle/nnQcompiler}{\url{https://github.com/Swaddle/nnQcompiler}}. This work was supported by resources provided by the Pawsey Supercomputing Centre with funding from the Australian Government and the Government of Western Australia \footnote{\url{https://www.pawsey.org.au/}}.

\bibliography{main.bib}
\appendix 
\section{Circuits for basis elements}
\label{section:circuit}
For each basis element, \( \tau_i\) in the Pauli basis for \( \mathfrak{su}(2^n) \), it is possible to construct a circuit for the exponential, \( \exp(\vartheta \tau_i)\) in \( \mathrm{SU}(2^n) \). The first step is to observe \( \exp( i \vartheta \sigma_3 \otimes \dots \otimes \sigma_3  ) \) has the following circuit representation
\begin{align*}
	\Qcircuit @C=0.4em @R=.4em{
		& \ctrl{1} & \qw &  \qw & \qw & \qw & \qw  &\qw  & \qw  &\qw    &  \qw   & \ctrl{1} &\\
		& \targ  & \ctrl{1} & \qw   &\qw  & \qw &\qw  &\qw & \qw  &  \qw & \ctrl{1} & \targ & \\
		&       &           &      &     &     &      &    &    &      &          &       &  \\
		&       & \vdots    &      &     &     &      &    &    &      & \vdots   &       & \\
		&       &  \vdots  &      &      &     &      &    &    &      & \vdots   &       & \\ 
    		& \qw   & \targ    &\ctrl{1} & \qw & \qw & \qw &\qw & \qw & \ctrl{1} & \targ & \qw & \\
    		& \qw & \qw   & \targ &  \ctrl{1} & \qw &  \qw                  &  \qw & \ctrl{1} & \targ  & \qw & \qw & \\ 
    		& \qw & \qw   & \qw   & \targ   & \qw   & \gate{R_3(\vartheta)} & \qw & \targ & \qw   & \qw  & \qw & \\}
  \end{align*}
where \( \vdots \) denotes staggered CNOT gates. When there is an identity matrix in the place of a \( \sigma_3 \), then the staggered CNOT gates skip the respective qubit. For example \( \exp(i \vartheta \sigma_3 \otimes I \otimes \sigma_3) \), has the circuit 
\begin{align*}
	\Qcircuit @C=0.4em @R=.4em {
    	& \qw &\qw & \ctrl{2} & \qw & \qw & \qw &  \qw & \qw & \ctrl{2}  & \qw  & \qw & \\ 
  	& \qw &\qw & \qw & \qw & \qw &  \qw &  \qw & \qw & \qw & \qw & \qw & \\ 
    & \qw & \qw & \targ & \qw  & \qw & \gate{R_3(\vartheta)} & \qw & \qw & \targ & \qw & \qw & \\
  	}
\end{align*}	
To obtain circuits for exponentials which contain \( \sigma_1 \) and \( \sigma_2 \) terms, make use of the \textit{Hadamard} gate, denoted \(H\), and the \(Y\) gate. They have the following matrix representations;
\begin{align*}
	H = \frac{1}{\sqrt{2}} \left( \begin{array}{cc}
		1 & 1 \\
		1 & -1 
	\end{array}\right),  \quad Y =\frac{1}{\sqrt{2}} \left(  \begin{array}{cc}
		1 & i \\
		i & 1 
	\end{array}\right),
\end{align*}	
and have the following properties
\[ Y \sigma_3 Y^{\dagger} = \sigma_2 , \quad H \sigma_3 H^{\dagger} = \sigma_1.\]
These gates can be used to swap the indices in the Kronecker product \( i \sigma_3 \otimes \dots \otimes \sigma_3 \). For example \( \exp(i \vartheta \sigma_1 \otimes I \otimes \sigma_2) \) has the circuit
\begin{align*}
	\Qcircuit @C=1em @R=.7em {
    	& \gate{H} &\qw & \ctrl{2} & \qw & \qw & \qw &  \qw & \qw & \ctrl{2}  & \qw  & \gate{H^{\dagger}}  & \\ 
  	& \qw &\qw & \qw & \qw & \qw &  \qw &  \qw & \qw & \qw & \qw & \qw & \\ 
    & \gate{Y} & \qw & \targ & \qw  & \qw & \gate{R_3(\vartheta)} & \qw & \qw & \targ & \qw & \gate{Y^{\dagger}} & \\
  	}
\end{align*}
This can be seen from the formula for the exponential of a single basis element. Since the circuit for 
\begin{align*} 
\exp(i \vartheta  \sigma_3 \otimes \dots \otimes \underbrace{\sigma_3}_{j-\text{th entry} } \otimes  \dots \otimes \sigma_3) 
\end{align*}
is known, to swap the \( j\)-th sigma, simply apply the \(H\) or \(Y\) gate to the \(j\)-th qubit. For example, to obtain a \( \sigma_1\) in the \( j\)th entry, let \( \widetilde{H} = I_{2\times2} \otimes \dots \otimes H \otimes  \dots \otimes I_{2\times 2}  \)
\begin{align*}		
&  \widetilde{H} \exp(  i \vartheta  \sigma_3 \otimes \dots \otimes \sigma_3  \otimes  \dots \otimes \sigma_3) \widetilde{H}  \\
&= \cos(\vartheta) I_{2^n \times 2^n} + i \sin(\vartheta) (\sigma_3 \otimes \dots \otimes \sigma_1  \otimes  \dots \otimes \sigma_3)\\
&= \exp( i \vartheta \sigma_3 \otimes \dots \otimes \sigma_1  \otimes  \dots \otimes \sigma_3 ),
 \end{align*}
which gives the desired result.

 \end{document}